\documentclass[12pt]{article}
\usepackage[dvipsnames]{xcolor}
\definecolor{MediumBlue}{rgb}{0.015, 0.315, 0.45}
\definecolor{cfOrange}{HTML}{f57900}
\usepackage{listings}
\usepackage{tikz}
\usetikzlibrary{calc,decorations.pathmorphing,shapes,arrows, shapes.arrows, positioning, shapes, fit,
matrix, decorations.pathreplacing,calligraphy,overlay-beamer-styles,backgrounds}
\tikzset{
  pobl/.style={
    inner sep=0pt, outer sep=0pt, fill=#1,
  },
  pobl gron/.style n args={2}{
    pobl=#1, rounded corners=#2,
  },
  pics/person/.style n args={3}{
    code={
      \node (-corff) [pobl gron={#1}{1pt}, minimum width=.25*#2, minimum height=.375*#2, rotate=#3, pic actions] {};
      \node (-pen) [minimum width=.3*#2, circle, pobl=#1, outer sep=.01*#2, anchor=south, rotate=#3, pic actions] at (-corff.north) {};
      \node (-coes dde) [pobl gron={#1}{1.5pt}, anchor=north west, minimum width=.115*#2, minimum height=.3*#2, rotate=#3, yshift=0.05*#2, pic actions] at (-corff.south west) {};
      \node (-coes chwith) [pobl gron={#1}{1.5pt}, anchor=north east, minimum width=.115*#2, minimum height=.3*#2, rotate=#3, yshift=0.05*#2, pic actions] at (-corff.south east) {};
      \node (-braich dde) [pobl gron={#1}{.75pt}, minimum width=.075*#2, minimum height=.325*#2, outer sep=.007*#2, anchor=north west, rotate=#3, pic actions] at (-corff.north east)  {};
      \node (-braich chwith) [pobl gron={#1}{.75pt}, minimum width=.075*#2, minimum height=.325*#2, outer sep=.007*#2, anchor=north east, rotate=#3, pic actions] at (-corff.north west) {};
    },
  },
}

\usepackage{float}
\usepackage[top=3cm, bottom=3cm, left=3cm, right=3cm, heightrounded,
  marginparwidth=2.6cm, marginparsep=3mm]{geometry}
\usepackage{adjustbox}
\usepackage{natbib}
\usepackage{chngcntr}
\definecolor{PlaceboBlue}{HTML}{6A97AD}

\usepackage{amsmath,amssymb}

\usepackage{amsmath}
\makeatletter
\let\save@mathaccent\mathaccent
\newcommand*\if@single[3]{%
  \setbox0\hbox{${\mathaccent"0362{#1}}^H$}%
  \setbox2\hbox{${\mathaccent"0362{\kern0pt#1}}^H$}%
  \ifdim\ht0=\ht2 #3\else #2\fi
  }
\newcommand*\rel@kern[1]{\kern#1\dimexpr\macc@kerna}
\newcommand*\widebar[1]{\@ifnextchar^{{\wide@bar{#1}{0}}}{\wide@bar{#1}{1}}}
\newcommand*\wide@bar[2]{\if@single{#1}{\wide@bar@{#1}{#2}{1}}{\wide@bar@{#1}{#2}{2}}}
\newcommand*\wide@bar@[3]{%
  \begingroup
  \def\mathaccent##1##2{%
    \let\mathaccent\save@mathaccent
    \if#32 \let\macc@nucleus\first@char \fi
    \setbox\z@\hbox{$\macc@style{\macc@nucleus}_{}$}%
    \setbox\tw@\hbox{$\macc@style{\macc@nucleus}{}_{}$}%
    \dimen@\wd\tw@
    \advance\dimen@-\wd\z@
    \divide\dimen@ 3
    \@tempdima\wd\tw@
    \advance\@tempdima-\scriptspace
    \divide\@tempdima 10
    \advance\dimen@-\@tempdima
    \ifdim\dimen@>\z@ \dimen@0pt\fi
    \rel@kern{0.6}\kern-\dimen@
    \if#31
      \overline{\rel@kern{-0.6}\kern\dimen@\macc@nucleus\rel@kern{0.4}\kern\dimen@}%
      \advance\dimen@0.4\dimexpr\macc@kerna
      \let\final@kern#2%
      \ifdim\dimen@<\z@ \let\final@kern1\fi
      \if\final@kern1 \kern-\dimen@\fi
    \else
      \overline{\rel@kern{-0.6}\kern\dimen@#1}%
    \fi
  }%
  \macc@depth\@ne
  \let\math@bgroup\@empty \let\math@egroup\macc@set@skewchar
  \mathsurround\z@ \frozen@everymath{\mathgroup\macc@group\relax}%
  \macc@set@skewchar\relax
  \let\mathaccentV\macc@nested@a
  \if#31
    \macc@nested@a\relax111{#1}%
  \else
    \def\gobble@till@marker##1\endmarker{}%
    \futurelet\first@char\gobble@till@marker#1\endmarker
    \ifcat\noexpand\first@char A\else
      \def\first@char{}%
    \fi
    \macc@nested@a\relax111{\first@char}%
  \fi
  \endgroup
}
\usepackage{amsthm}
\usepackage{pifont}

\usepackage{textcomp}
\usepackage{sansmath}
\usepackage{tcolorbox}
\usepackage{appendix}
\usepackage{soul}
\sethlcolor{red!60}

\usepackage{pdflscape}
\usepackage[labelfont=bf]{caption}
\usepackage{subcaption}
\theoremstyle{definition}
\newtheorem{definition}{Definition}

\newtheorem{innercustomgeneric}{\customgenericname}
\providecommand{\customgenericname}{}
\newcommand{\newcustomtheorem}[2]{%
  \newenvironment{#1}[1]
  {%
   \renewcommand\customgenericname{#2}%
   \renewcommand\theinnercustomgeneric{##1}%
   \innercustomgeneric
  }
  {\endinnercustomgeneric}
}
\newcustomtheorem{assumption}{Assumption}

\usepackage{mathtools}
\usepackage{bm}
\usepackage{framed}
\usepackage{longtable}
\usepackage{ragged2e}
\usepackage{makecell}

\usepackage{longtable}
\usepackage{tabularx,ragged2e}
\newcolumntype{C}{>{\Centering\arraybackslash}X}
\newcolumntype{R}{>{\raggedleft\arraybackslash}X}
\newcolumntype{L}{>{\raggedright\arraybackslash}X}
\usepackage[roman, breakwithin]{parnotes}
\usepackage{array}
\usepackage{array}

\usepackage{graphicx}
\usepackage{setspace}

\makeatletter
\newcommand{\smallsym}[2]{#1{\mathpalette\make@small@sym{#2}}}
\newcommand{\make@small@sym}[2]{%
  \vcenter{\hbox{$\m@th\downgrade@style#1#2$}}%
}
\newcommand{\downgrade@style}[1]{%
  \ifx#1\displaystyle\scriptstyle\else
    \ifx#1\textstyle\scriptstyle\else
      \scriptscriptstyle
  \fi\fi
}
\makeatother
\usepackage{hyperref}

\usepackage{appendix}

\title{Notes on Randomized Controlled Trials for Studying Social Media Harms\thanks{This is a restructured and extended version of an earlier paper that circulated under the title ``Why Small Experimental Effects of Social Media Use are Compatible with Large Real-World Effects.'' For comments on earlier drafts, I am grateful to Simone Zhang, Tyler VanderWeele, Brandon Stewart, Luke Miratrix, Ian Lundberg, Jon Haidt, and Han Choi. For helpful discussions related to the project, I thank Mara Ishihara Zinky, Noah Padgett, Dalton Conley, members of the Miratrix CARES Lab, and members of Jon Haidt's social media lab. I owe special debts to Zach Rausch and Felix Elwert: Zach provided extremely thorough feedback on multiple drafts, and Felix both encouraged me to restructure the paper and offered invaluable guidance for doing so. Errors are mine.}}
\author{Chris Felton\thanks{Post-doctoral Fellow, Institute of Religious Studies, Baylor University, and Visiting Scholar, Human Flourishing Program, Harvard University. \href{mailto:cmfeltonwork@gmail.com}{\texttt{cmfeltonwork@gmail.com}} will forward to current and future academic e-mail addresses.}}
\date{March 23, 2026}

\begin{document}
\maketitle

\begin{abstract}

Randomized controlled trials (RCTs) and person-level observational studies feature prominently in debates over social media harms. I highlight some under-acknowledged limitations of such evidence. Most important is that published RCTs typically identify effects of a \textit{local}, or small-scale, intervention: a person is assigned to quit social media, but her immediate peers continue using it in large numbers. Critics of social media, in contrast, focus on a \textit{global}, or large-scale, intervention: the mass adoption of social media among U.S. teenagers. Such global interventions alter both the proximal social environment and the broader culture, potentially harming teenagers who abstain from social media entirely. This paper discusses the local--global distinction at length and offers other notes on the limits of learning about social media harms from existing RCTs and person-level observational studies. I suggest that triangulating different forms of imperfect evidence may provide the deepest insights about social media's aggregate effect on teen mental health.

\end{abstract}

\newpage

\section{Introduction}\label{sec:intro}

Since 2012, the mental health of US teenagers has deteriorated. Suicide and self-harm rates have risen along with self-reports of depression and psychological distress \citep{trends1,trends2,choi}.\footnote{The rise in teen suicide mortality starts around 2008 rather than 2012; see \citet{smh2} for further discussion. See also \citet{choi}, who note that the rise in emergency department visits for self-harm remains substantial even after accounting for coding changes.}  One explanation points to widespread smartphone and social media use among teenagers as a major contributor to these trends \citep{smh1,smh2}.\footnote{\citet{smh2} additionally suggests play deprivation and overprotection play important roles as well.} To ease exposition, I focus specifically on social media use and refer to this explanation as the \textit{social media hypothesis}.\footnote{I discuss smartphones further in \autoref{sec:compound}. The \textit{social media hypothesis} is not a standard term.} 

Critics of the social media hypothesis point to evidence that, on the surface, appears to contradict the explanation \citep{small1,small2,small3,small4,small5,small6,small7}. Many randomized controlled trials (RCTs), according to critics, show that encouraging heavy social media users to quit has only small effects on mental health in the short term \citep{exp1,exp2,exp3,exp4,exp5,exp6,exp7,exp8,exp9,exp10,exp11,exp12,burnell}.\footnote{The characterization of some of the RCT estimates as ``small'' is contested \citep{group2}. I take no stance on this issue in the present paper.} Cross-sectional associations between social media use and mental health appear to be similarly small \citep{OP}. Were social media the primary culprit for the stark decline in mental health, the argument goes, effects would be much larger. Factors other than social media must explain the majority of this drop. 

In this paper, I highlight an important problem with this argument: extant social media RCTs fail to offer an appropriate test of the social media hypothesis. Were we to design an ideal RCT---one that provided the best opportunity to assess the explanation---it would differ drastically from the RCTs that have actually been conducted. This paper focuses on three differences in particular. 

First, the type of intervention in the ideal trial would differ. The social media hypothesis concerns the effects of joining social media; published RCTs typically identify the effect of quitting social media, often after years of use. Second, the intervention's duration would differ. The social media hypothesis concerns the years-long exposure to social media; RCTs typically identify the effect of quitting social media for only several weeks. Finally, and perhaps most importantly, the intervention's scale would differ in this ideal trial. The social media hypothesis concerns a \textit{global}, or large-scale, intervention: the mass adoption of social media among U.S. teenagers. RCTs, in contrast, identify effects of a \textit{local}, or small-scale, intervention whereby a teenager quits social media but the vast majority of her peers continue using it. Such local and global interventions may diverge substantially in their effects on mental health, as RCT participants who quit social media may still be affected by the social media habits of non-participants. When social media use is ubiquitous, it can fundamentally alter both the proximal social environment and the broader culture, affecting the mental health of teenagers who abstain from social media entirely. In short, existing RCT estimates---even if we agree to characterize them as small---provide little evidence against the social media hypothesis. 

I thus aim to contribute to an ongoing debate about social media harms. Social media is a contentious topic, however, and I should emphasize that this paper does not offer a wholesale endorsement of the social media hypothesis. Economic factors have plausibly harmed teen mental health \citep{econ1,econ2}, and accurately quantifying how much of the total decline in mental health can be attributed to different causes is fraught with methodological challenges (see \autoref{sec:attribution}). The substantive contribution of the paper is simply to defend the social media hypothesis against one particular strain of criticism. 

I also aim to illustrate a few general methodological points. The first is that when studying phenomena that potentially transform social environments, cluster-randomized trials---as well as cluster-level observational studies---can yield critical insights that individually-randomized trials cannot. An individually-randomized trial enables a researcher to identify effects of a small-scale intervention that typically leaves the proximal social environment unaltered. Cluster-randomized trials, in contrast, can credibly identify the effects of large-scale interventions that alter group dynamics and social networks. Individually-randomized social media trials are still valuable: their estimates can inform a person's decision to quit social media or a parent's decision to allow their child to join. But if we wish to learn more about the social media hypothesis---or public policies aimed at restricting social media use---cluster-randomized trials can better serve us. Cluster-randomized trials may also prove to be invaluable tools for studying other global interventions like mass exposure to misinformation or widespread access to generative artificial intelligence. 

Two additional methodological points concern recent calls for social scientists to more precisely define estimands of interest \citep{hernanC,rohrer1,lundberg,rohrer2}. (An \textit{estimand} is the true quantity an analyst seeks to estimate. One example estimand might be the causal effect of quitting social media for one week among all heavy social media users in the US. A one-week social media trial can provide a single \textit{estimate} of this quantity, but it will differ from the true value of the estimand due to sampling variability. Estimands are also sometimes called \textit{target quantities} or \textit{population parameters}.)

These scholars have rightly condemned the ``taboo against explicit causal inference'' in observational studies, arguing that this convention weakens research design and complicates the interpretation of results. I wish to add that precisely specifying causal estimands is just as important in experimental research. Much of the disagreement over social media harms stems from the often unstated differences between the estimands most credibly identified by RCTs and the estimands most relevant for evaluating the social media hypothesis. Clarifying these differences can inform future work on social media and help resolve points of disagreement between scholars.

Finally, I highlight that specifying hypothetical randomized trials offers a useful way for researchers to precisely define different causal estimands.\footnote{Specifying (and then emulating) target trials can also help researchers design better observational studies \citep{dorn1953,cochran1965,hernan2016}. This paper focuses on the use of hypothetical trials for defining causal estimands \citep{haavelmo,hernan2005}.} I am far from the first to make this point, but target trial specification is usually discussed in the context of observational research \citep{haavelmo,hernan2005}. Contrasting an ideal trial with an actual trial can help clarify what the latter teaches us about the social world. Moreover, target trial specification remains rare even in observational studies. My hope is that an extended discussion of target trial specification in the context of the social media debate will illustrate the benefits of this exercise to a wider audience of social scientists.

\subsection{Related Work and Contribution}

A large body of work explores the potential harms of social media. Most relevant to this paper is the debate over whether the magnitude of social media's effect on mental health---as estimated in individually-randomized trials and person-level observational studies---supports the social media hypothesis \citep{OP,twengeOP,group2}. Much of this debate focuses on whether these published effect sizes should be considered ``small,'' or ``policy-irrelevant.'' I take no stance on this issue in the present paper and instead concentrate exclusively on what types of effects are being estimated in the first place. 

This paper is not the first to highlight the importance of distinguishing between different estimands in social media research. For instance, some have argued that average treatment effects obscure effect heterogeneity by gender, with social media apparently harming girls more severely than boys \citep{twengeOP,group2}.\footnote{Other sources of heterogeneity include age, platform, trans identity, and pre-existing mental health issues \citep{het,group2,het1,het2,het3}.} But the specific points I detail in this paper have been largely overlooked. Intervention scale in particular has received relatively little attention. A notable exception is \citet{collectiveaction}, who offer the richest empirical study related to global social media interventions to date, although they focus on one specific consequence of global interventions---namely, social exclusion stemming from collective-action traps in social media use. \citet{group1} and \citet{group2} have discussed the issue using the term ``group-level effects'' but also primarily focus on social exclusion (\citet{group1}, p. 22; \citet{group2}, p. 38).\footnote{I explain why I avoid the term ``group-level effects'' in \autoref{fn:group}.}

This paper expands substantially on this prior work in several ways. First, I describe several different types of spillover that may cause local- and global-intervention effects to differ: the sharing of social media content with non-users, the contagion of mental health outcomes, the aforementioned social exclusion phenomenon, and the adaptation of larger institutions to widespread social media use. Second, I offer formal definitions of local- and global-intervention effects, grounding discussions of group-level effects in the potential outcomes framework and relating it to the causal inference literature on interference \citep[e.g.,][]{hudgens,vand1}. These definitions both illustrate why global interventions have many effects beyond social exclusion and clarify that drawing inferences about the social media hypothesis based on extant RCTs relies on strong and typically unstated assumptions. Finally, I suggest ways for social scientists to empirically assess whether global social media interventions have appreciably larger effects than their local counterparts. These suggestions can inform future work on social media harms and the importance of intervention scale.

This paper also draws on methodological work from several disciplines. Economists recognize the importance of intervention scale in the distinction between \textit{partial-} and \textit{general-equilibrium} effects \citep{heckman1}. Because markets are interactive, the proportion of market participants exposed to a particular treatment strongly influences aggregate outcomes. Epidemiologists consider intervention scale in vaccine studies \citep{hudgens,vand1,vand2}. Infections travel from person to person, so one person's catching a virus is partly a function of how many of her peers are vaccinated. In statistics, a growing literature focuses on identifying effects of global interventions using cluster-randomized trials \citep{molly,leung1,leung2,linear}. And in sociology, scholars have devised both observational and experimental methods for bringing causality into the study of social networks \citep{networkexp,duxbury1,duxbury2}. This paper applies these methodological insights to the ongoing debate about social media harms, partly to contribute to the debate itself, and partly to illustrate the importance of intervention scale for a more applied audience. 

Finally, the paper is related to work on the benefits and drawbacks of RCTs \citep{imbens,sampson,deaton,rocio}. I take a less critical stance toward RCTs than \citet{sampson} and \citet{deaton} but emphasize the limits of learning from individually-randomized trials on social media harms. Even cluster-randomized trials, which may be more useful than individually-randomized trials, will suffer from various shortcomings. Ultimately, I suggest that we can learn the most by triangulating different sources of evidence in the spirit of \citet{cornfield}.\footnote{See also \citet{haack} and \citet{levimartin}.}

\subsection{Structure of the Paper}

The paper is structured as follows. \autoref{sec:accessible} illustrates the differences between the RCT estimand and the social media hypothesis estimand using hypothetical randomized trials. This section devotes special attention to the distinction between local and global interventions and why their per-person effects may differ in magnitude. \autoref{sec:technical} offers a more formal discussion of local and global interventions as well as interference assumptions in causal inference. \autoref{sec:further} offers additional notes on causal estimands related to the social media hypothesis as well as a brief remark on cross-sectional observational studies on social media harms. \autoref{sec:conclusion} concludes with thoughts on how we might learn more about the effects of global interventions when identification assumptions fail. 

\section{Causal Estimands in Social Media RCTs: A Non-Technical Discussion}\label{sec:accessible}

This section highlights three core differences between the RCT estimand and the social media hypothesis estimand: intervention type, intervention duration, and intervention scale. Intervention type and intervention duration are closely related in this setting, but I discuss them separately to ease exposition.

Following \citet{hernan2005} I use hypothetical randomized trials to precisely characterize different causal estimands without formal notation (see also \citet{haavelmo}). For the sake of illustration, these hypothetical trials are somewhat idealized, but I attempt to idealize the trials in a way that is favorable to critics of the social media hypothesis. For instance, many social media RCTs encourage participants to merely reduce time spent on social media, but I describe trials that encourage participants to quit social media entirely. Footnotes will describe such discrepancies in more detail. I also assume away issues like attrition and non-compliance, again for the sake of illustration. \autoref{tab:three} summarizes the main points of the section.
    
\subsection{Intervention Type: Joining vs. Quitting}\label{sec:type}

Most social media RCTs identify the effects of quitting social media, most likely after years of use.\footnote{This idealized trial differs from real-world social media trials in a few ways. First, many social media RCTs examine reduced use rather than quitting. Second, many RCTs assign participants to quit only one social media platform rather than all social media platforms. Third, RCT participants are typically adults, not teenagers, and using social media as a young teenager may have longer-lasting consequences \citep{het}. Finally, RCTs typically do not require a specific period of prior use and the published papers usually do not report the duration of prior use. I specify two years of prior use for \textit{Quitting Trial}, which is plausible as an average duration of prior use for 16-year-old users; research in the United States and United Kingdom suggests a majority of adolescents join social media before age 13 \citep{prevalence1,prevalence2}.} But the social media hypothesis concerns the effects of joining social media for the first time. To appreciate the difference, consider the following hypothetical trials. 

\newpage

\begin{quotation}

\noindent\textbf{Joining Trial.} Researchers recruit a sample of teenagers who have never used social media. Never-users are either encouraged to either join social media (treated) or continue abstaining (control).\footnote{A real trial would not encourage participants to join social media but instead encourage them to remain off social media. I describe a trial in which researchers encourage joining to facilitate exposition.}

\end{quotation}

\begin{quotation}

\noindent\textbf{Quitting Trial.} Researchers recruit a sample of teenagers who have used social media daily for the past two years. Users are randomly assigned either to stop using social media (treated) or to continue using social media (control). 

\end{quotation}

\noindent In each trial, researchers measure depressive affect scores after three weeks and estimate the treatment effect as the difference between the mean scores of treated and control participants. 

The estimated effects from these trials will likely differ in direction, with joining being harmful and quitting beneficial. But the effects may also differ in magnitude, with joining producing larger absolute effects than quitting. Suppose Alice, who participates in \textit{Quitting Trial}, started using social media heavily at a young age.\footnote{See the discussion of average joining age in previous footnote.} Suppose further that this led her to develop a persistent social comparison habit: she frequently compares herself with both peers and ``influencers'' in a way that damages her mental health \citep{compare1,compare2}. In compliance with her assigned treatment, she quits social media for three weeks after using it for two years. Quitting at this stage, however, may do little to curb her social comparison tendencies. In contrast, suppose Bill participates in \textit{Joining Trial} and starts using social media for the first time. Three weeks of heavy use at an impressionable age may be enough for him to develop a social comparison habit similar to Alice's, at least in the short term. Of the two trials, \textit{Joining Trial} comes closer to the ideal trial we might run to assess the social media hypothesis.

More generally, what is notable about joining and quitting trials is that they require distinct, non-overlapping populations: a quitting trial cannot be run on never-users, and a joining trial cannot be run on current users. Borrowing terminology from the causal inference literature, the two populations are defined by different \textit{treatment histories} \citep{hernan}.\footnote{I am slightly abusing terminology: \textit{treatment history} usually describes a participant's history of treatment within the study period, not before the study period.} For instance, suppose that all trial participants are 16 years old and that \textit{Quitting Trial} participants have used social media for the past two years but no more. Each participant would have a treatment history of abstaining from social media for 14 years followed by using social media for two years. Participants in \textit{Joining Trial}, in contrast, would have treatment histories of 16 years of abstention at baseline. Assuming the quitting estimand and joining estimand are equal in magnitude amounts to assuming that the past two years of a participant's treatment history do not moderate the effect of using social media for the next three weeks. The concept of treatment history will be of further use in discussing intervention duration, which I turn to next.

\subsection{Intervention Duration: Short vs. Long}\label{sec:duration}

Social media RCTs typically target the effects of quitting social media for between one and three weeks \citep{exp3,exp4,exp5,exp6}. To appreciate the importance of intervention duration, consider two new hypothetical trials. The first is identical to \textit{Quitting Trial}: participants either quit or continue using social media for three weeks. In the second trial, participants either quit or continue using social media for three years. 

The two effects plausibly differ in magnitude. The benefits of cessation may increase with its duration: the longer someone avoids social media, the more her mental health may improve. Three years' abstinence may do more to improve Alice's social comparison habit than three weeks' abstinence. Of course, it is also possible that benefits will subside after some time, consistent with psychological theories of hedonic adaptation \citep{hedonic}. Perhaps the novelty of quitting social media only temporarily improves Alice's mood, after which she returns to her baseline level of well-being. The point is that the estimands differ and that an ideal trial would capture effects of long- rather than short-term use.

Both intervention duration and intervention type, as I have described them, are related to treatment history. In particular, intervention type is tied to the \textit{pre}-baseline treatment histories of different trial participants, whereas intervention duration concerns \textit{post}-baseline treatment histories. The two issues can be captured simultaneously by specifying two distinct pairs of treatment histories corresponding to two different hypothetical trials. In an ideal hypothetical RCT designed to learn about the social media hypothesis, we would compare the following pair of treatment histories:

\begin{enumerate}

    \item[(i)] 14 years of abstention, followed by two years of social media use; and 
    \item[(ii)] 16 years of abstention.

\end{enumerate}

\noindent This ideal trial would effectively be a longer-running version of \textit{Joining Trial} with participants sampled at a younger age. In real-world RCTs, however, we typically compare treatment histories similar to the following:

\begin{enumerate}

    \item[(i)] 14 years of abstention, followed by two years and three weeks of social media use; and
    \item[(ii)] 14 years of abstention, followed by two years of social media use, followed by three weeks of abstention. 

\end{enumerate}

\noindent Even before taking intervention scale into consideration, it is unlikely these two comparisons produce effect sizes of equal magnitude.

\begin{landscape}

\begin{table*}[t]
\captionsetup{font = footnotesize}
\vspace{2cm}
{\renewcommand{\arraystretch}{1.5}%
\begin{tabularx}{\linewidth}{l X  X  X }
\hline
&\textbf{Social Media RCTs} & \textbf{The Social Media \newline Hypothesis} & \textbf{When might the difference matter?}  \\
\hline

\textit{Intervention Type} & Quitting social media after years of use & Joining social media for the first time & When joining social media has long-lasting effects not undone by quitting \\

\noalign{\vskip 2mm}    

\textit{Intervention Duration} & Short (often 1--3 weeks)  & Long (several years) & When long-term use of (or abstention from) social media has larger effects than short-term use (or abstention) \\

\noalign{\vskip 2mm}    

\textit{Intervention Scale} & Local (negligible fractions of peer group and broader population receive treatment) & Global (large fractions of peer group and broader population receive treatment) & When widespread adoption of social media affects those who abstain from it, such as by altering the proximal social environment or broader culture \\

\hline

\end{tabularx}}
\caption{\textbf{Three ways that the estimand identified by RCTs differs from the estimand invoked by the social media hypothesis.} Note that the table focuses on typical, existing RCTs related to social media use; different RCTs could identify different types of causal effects.}
\label{tab:three}
\end{table*}

\end{landscape}

\subsection{Intervention Scale: Local vs. Global}\label{sec:scale}

Social media RCTs identify the effects of local interventions, whereas the social media hypothesis concerns the effect of a global intervention. To illustrate the difference between local- and global-intervention effects, suppose that researchers have randomly selected a set of high schools from the population. Assume that a high school student's mental health can be affected by the social media habits of other students in her school but not by the social media habits of students in other schools (or of anyone else in the population). Suppose that in each school, the majority of students use social media. For simplicity, we will also suppose each high school has the same number of students. Now consider two potential trials researchers might run.

\begin{quotation}

\noindent\textbf{Local Quitting Trial.} Researchers randomly select one social media user from each school. The student is assigned either to quit social media or to continue using it, and researchers do not intervene on any of her peers. Three weeks later, depressive affect scores are measured for the selected students only. The estimated treatment effect of local quitting is the difference between the mean scores of the selected treated and control students.\footnote{See \autoref{sec:real} for a discussion of real-world individually-randomized trials.}

\end{quotation}

\begin{quotation}

\noindent\textbf{Global Quitting Trial.} Researchers sample every student from each school. In treated schools, every student is assigned to quit social media. In control schools, nobody is encouraged to quit social media. Three weeks later, depressive affect scores are measured for all students. The estimated treatment effect of global quitting is the difference between the mean scores of treated and control students.

\end{quotation}

\noindent To see why these effects might differ, consider how things would change for Alice between the two trials, assuming she would be assigned to the treatment condition in either case. In \textit{Local Quitting Trial}, she would remain partly exposed to the harms of social media through her peers' routine use---a phenomenon known as \textit{spillover} or \textit{interference}.\footnote{Note that in RCTs, the finite-population formulation of the no-interference assumption prohibits interference between trial participants only. Consequently, this version of the assumption is not necessarily violated in individually-randomized trials. The super-population version of the assumption, however, prohibits interference between any two members of the broader population, which almost certainly fails to hold. See \autoref{sec:interference} for further discussion.} In \textit{Global Quitting Trial}, however, she would be totally unexposed to the harms of social media, save for the longer-term effects of prior use.

An analogy may be drawn with the effects of quitting smoking. (I hasten to add that the purpose of the analogy is to help illustrate the difference between local- and global-interventions---not to imply that social media is as harmful as smoking or that the evidence of harm is comparable in volume or quality.) Imagine a cramped, poorly ventilated apartment building filled with heavy smokers. A local intervention would entail encouraging only one person to quit smoking, whereas the global intervention would entail encouraging everybody to quit. It is easy to see why the local intervention would produce smaller effects on lung cancer prevalence than would the global intervention: under the local intervention, the excessive second-hand smoke continues to harm the sole quitter. In an individually-randomized social media trial, excessive social media use by peers may likewise continue to harm the sole member of a social network who quits. The following subsection will describe how this might occur.
                
\subsubsection{Potential Sources of Difference Between Local- and Global-Intervention Effects}\label{sec:why}

In what follows, I describe different examples of spillover effects that plausibly lead local interventions to have smaller effects than global interventions, focusing specifically on quitting interventions. This is not an exhaustive list of examples, and I describe other potential spillover effects in footnotes. I also depict some of these spillover effects graphically in \autoref{fig:notDAG}. 

\textit{Content Sharing.} One potential spillover effect occurs through sharing social media content with non-users. Suppose Alice's best friends are heavy social media users and that their routine use has caused them to develop harmful social comparison habits. Suppose further that Alice enrolls in an RCT and is assigned to quit social media. She complies with the assigned treatment, but her friends, who are not enrolled in the trial, continue using social media. When Alice spends time with her friends, they often show her social media posts shared by schoolmates, inducing continued social comparison behavior in Alice. As a result, the local intervention has limited effects on her mental health: despite personally abstaining from social media, she remains partly exposed to social media content, leading to only modest gains in well-being. 

\textit{Mental Health Contagion.} Another reason local- and global-intervention effects might differ is that mental health can be contagious \citep{contagion1,contagion2,contagion3,contagion4,contagion5}. Suppose Alice's friends use social media regularly and that this routine use increases anxiety and depressive symptoms. If such symptoms are contagious---perhaps through the spread of thinking styles or co-rumination---her quitting may again have only limited effects. Even if she avoids social media content entirely, she will remain exposed to social media--induced changes in her peers' behavior. 

\textit{Social Exclusion and Isolation.} Peer behavior may affect Alice through another pathway: social exclusion and isolation brought on by widespread social media use.\footnote{Content sharing, mental health contagion, and social exclusion are three examples of how social media--induced changes in peer behavior may affect a non-user, but there are others. For instance, suppose Alice's friends browse social media while spending time with her. This behavior may reduce the ``quality'' of the time they spend together even if her friends avoid sharing any social media content with her, which may in turn have downstream effects on Alice's mental health. Such spillover would also contribute to the difference between local- and global-intervention effects.} Suppose Alice's friends use social media to the extent that they rarely socialize offline. Alice stops using social media in compliance with her assigned treatment status and consequently consumes no social media content. But she is now much more socially excluded than before. If her friends primarily use social media to socialize with one another, abstaining might make her feel left out. Moreover, her friends' excessive use might leave her with fewer opportunities to socialize in person even if she does not feel left out. Alice, then, faces a lose--lose situation: if she rejoins social media, she will be directly exposed to its harms but less socially isolated; if she continues abstaining, she will avoid the harms stemming from direct exposure but remain isolated from her peers. This is one form of spillover that proponents of the social media hypothesis have highlighted \citep{group1,group2}. \citet{collectiveaction} presents compelling evidence that these social media ``collective-action traps'' exist.

\textit{Institutional Adaptation.} A final reason local- and global-intervention effects may differ is that institutions change in response to widespread social media use. Suppose that negative news stories are more widely shared on social media than positive stories, leading news outlets publish more of the former \citep[see, e.g.,][]{neg1,neg4,neg5}. If Alice is an avid news reader, these publishing practices may continue to harm her mental health after she quits social media \citep{neg1eff,neg2eff,neg3eff,neg4eff}. Other forms of institutional adaptation, however, may improve mental health. For instance, widespread social media use may pressure social media companies to remove harmful content or increase demand for products that prevent excessive social media use, such as time-limit apps or lockboxes for smartphones. 

Effects of social media use that operate through institutional adaptation are difficult to capture in real-world RCTs and were assumed away in \autoref{sec:scale} for illustration. Capturing such effects would require substantial portions of an entire country's population to quit social media. If these indirect effects are small relative to the indirect effects of peer usage, however, cluster-randomized trials may come close to identifying global-intervention effects.

\subsubsection{The Ideal RCT and Real-World RCTs}\label{sec:real}

The ideal trial for assessing the social media hypothesis would institute a long-term global joining intervention. This might proceed by encouraging entire schools of never-users to remain off social media while providing no such encouragement to other schools. In the absence of both non-compliance and between-school interference, such a trial would offer a more appropriate test of the social media hypothesis than do existing trials.

But real-world RCTs are more complicated. Interference across schools may occur, and compliance would likely be low. Between-school interference would threaten the identification of global-intervention effects, and low compliance would render the estimand less scientifically interesting, as the social media hypothesis concerns actual social media use rather than encouragement to use social media. Real-world cluster-randomized trials may therefore serve as a useful but imperfect tool for assessing the social media hypothesis (see \autoref{sec:conclusion} for further discussion).

\begin{landscape}
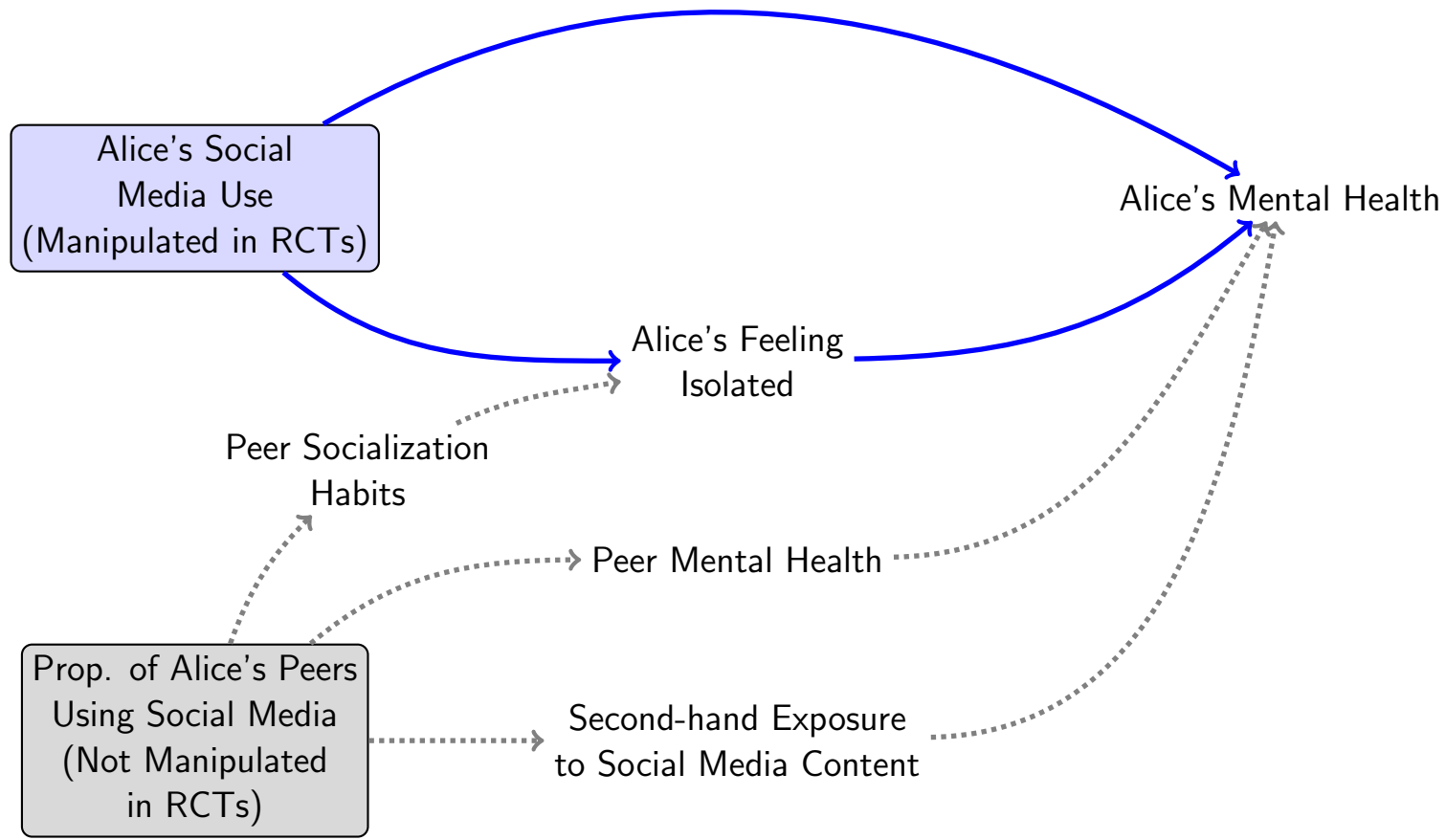
\begin{figure}
\centering
\begin{tikzpicture}[x = 7.5cm, y = 5cm, font=\large\sffamily]
\node[draw, rectangle, thick, inner sep = 4pt, rounded corners, fill = blue!15] (aliceUse) at (-.2, 0) [align=center]{Alice's Social \\ Media Use \\ (Manipulated in RCTs)};
\node (aliceMH) at (1.8,0) {Alice's Mental Health};
\node (aliceIso) at (.8,-.45) [align=center]{Alice's Feeling \\ Isolated};
\node[draw, rectangle, thick, inner sep = 4pt, rounded corners, fill = gray!30] (friendsUse) at (-.2, -1.5) [align = center]{Prop. of Alice's Peers \\ Using Social Media \\ (Not Manipulated \\ in RCTs)};
\node (pMH) at (.8, -1) [align = center]{Peer Mental Health};
\node (SH) at (.8, -1.5) [align = center]{Second-hand Exposure \\ to Social Media Content};
\node (pS) at (.1, -.75) [align = center]{Peer Socialization \\ Habits};

\draw[->, line width=2pt, blue] (aliceUse) to[out = 30, in = 150] (aliceMH);
\draw[->, line width=2pt, blue] (aliceUse) to[out = -40, in = 180] (aliceIso);
\draw[->, line width=2pt, blue] (aliceIso) to[out = 1, in = -140] (aliceMH);
\draw[->, line width=2pt, gray, dotted] (friendsUse) to[out = 70, in = -135] (pS);
\draw[->, line width=2pt, gray, dotted] (pS) to[out = 25, in = -170] (aliceIso);
\draw[->, line width=2pt, gray, dotted] (friendsUse) to[out = 40, in = -180] (pMH);
\draw[->, line width=2pt, gray, dotted] (pMH) to[out = 1, in = -120] (aliceMH);
\draw[->, line width=2pt, gray, dotted] (friendsUse) -- (SH);
\draw[->, line width=2pt, gray, dotted] (SH) to[out = 1, in = -100] (aliceMH);

\end{tikzpicture}
\caption{\small\textbf{Published RCTs typically fail to capture important causal effects stemming from widespread adoption of social media.} They primarily capture the effects represented by blue, solid arrows and mostly fail to capture the effects represented by gray, dashed arrows. (If Alice's treatment status affects her peers' social media use, RCTs may partly capture those effects.) To simplify the figure, I ignore the proportion of the broader population using social media, which may further affect Alice's mental health through institutional adaptation (see \autoref{sec:why}).}
\label{fig:notDAG}
\end{figure}
\end{landscape}

It should also be noted that existing individually-randomized trials vary in how well they may approximate the effects of what I have called a local-intervention, in which only a single member of a larger social network or community is assigned to quit social media. \citet{exp5}, for instance, recruit 80 Italian-speaking Instagram users via social media advertisements. It is possible some trial participants knew each other, but the total number of social ties is likely small. Similarly, \citet{exp3} recruit 180 participants through a combination of social media posts and invitations displayed at public places in Germany. These trials plausibly come close to identifying local-intervention effects. In contrast, \citet{exp6} select a sample of students from the same university, as do \citet{exp7}. In such studies, many trial participants may know one another, generating substantial within-sample interference. These trials may still identify what \citet{eate} term the \textit{expected average treatment effect}. Although this quantity differs from a pure local-intervention effect, it is similar in that it is an average of ``direct'' effects of changing only the focal participant's treatment status.\footnote{The average local-intervention effect differs because the treatment assignments for everybody but the non-focal person are fixed at the control value, whereas ``direct effects'' more generally fix the treatment assignments for everybody but the non-focal person but not necessarily at the control value.} Consequently, it still represents the average effect of a small-scale intervention. 

\section{Causal Estimands in Social Media RCTs: Technical Details and Nuances}\label{sec:technical}

This section provides a more formal basis for some of the points raised in the preceding discussion. I first define different types of local- and global-intervention effects in the potential outcomes framework. I then remark on a subtlety concerning no-interference assumptions in causal inference.

Before proceeding, I note a departure from convention. Typically, causal estimands are defined either in the \textit{super-population} framework or in the \textit{finite-population} framework. In the former, we consider the sample to be drawn from a hypothetical, infinitely large population. In the latter, the sample constitutes the entire population of interest. These frameworks accommodate statistical inference more easily than would a framework in which we draw the sample from a large but finite population. Because I do not consider statistical inference in this paper, however, I instead follow \citet{imai2008} in representing samples as draws from finite populations. This sample-from-a-finite-population framework allows us to distinguish between members of the population who participate in the trial and those who do not, which will prove useful in discussing interference.

\subsection{Formal Definitions of Local- and Global-Intervention Effects}\label{sec:estimands}

Suppose researchers wish to learn about a large, finite population comprising members indexed by $i = 1,2,\ldots,n$. For illustration, suppose this is the population of all high school students in a country and that a student's mental health is unaffected by the social media habits of anybody who is not a high school student. $Y_i$ represents person $i$'s depressive affect score, and $Z_i$ represents whether researchers encouraged her to quit social media ($Z_i = 1$). $Y_i(Z_i = 1)$ represents her depressive affect score that researchers would observe if, possibly contrary to fact, she had been encouraged to quit social media---that is, her potential outcome under treatment \citep{rubin1974}. $Y_i(Z_i = 0)$, in contrast, represents her potential outcome under control. Person $i$'s individual treatment effect is defined as $Y_i(Z_i = 1) - Y_i(Z_i = 0)$, and the population average treatment effect is defined as $\frac{1}{n}\sum_{i=1}^n Y_i(Z_i = 1) - Y_i(Z_i = 0)$. The fundamental problem of causal inference is that we can never observe both $Y_i(Z_i = 1)$ and $Y_i(Z_i = 0)$ for the same person $i$ \citep{holland1986}. The population average treatment effect can therefore only be estimated---not computed exactly---even if we could collect data on every single member of the population.

\begin{figure}
\centering
\resizebox{.75\textwidth}{!}{
\begin{tikzpicture}
\matrix (m1) [
  matrix of nodes,
  nodes={inner sep=0pt},
  column sep=15pt,
  row sep=6pt,
  ampersand replacement=\&
] {
  \tikz\pic{person={MediumBlue}{45pt}{0}}; \&
  \tikz\pic{person={MediumBlue}{45pt}{0}}; \&
  \tikz\pic{person={MediumBlue}{45pt}{0}}; \\
  \tikz\pic{person={MediumBlue}{45pt}{0}}; \&
  \tikz\pic{person={MediumBlue}{45pt}{0}}; \&
  \tikz\pic{person={MediumBlue}{45pt}{0}}; \\
  \tikz\pic{person={cfOrange}{45pt}{0}}; \&
  \tikz\pic{person={MediumBlue}{45pt}{0}}; \&
  \tikz\pic{person={MediumBlue}{45pt}{0}}; \\
};

\matrix (m2) [
  matrix of nodes,
  nodes={inner sep=0pt},
  column sep=15pt,
  row sep=6pt,
  ampersand replacement=\&,
  right=45pt of m1.east
] {
  \tikz\pic{person={MediumBlue}{45pt}{0}}; \&
  \tikz\pic{person={MediumBlue}{45pt}{0}}; \&
  \tikz\pic{person={MediumBlue}{45pt}{0}}; \\
  \tikz\pic{person={MediumBlue}{45pt}{0}}; \&
  \tikz\pic{person={MediumBlue}{45pt}{0}}; \&
  \tikz\pic{person={MediumBlue}{45pt}{0}}; \\
  \tikz\pic{person={MediumBlue}{45pt}{0}}; \&
  \tikz\pic{person={MediumBlue}{45pt}{0}}; \&
  \tikz\pic{person={MediumBlue}{45pt}{0}}; \\
};

\matrix (m3) [
  matrix of nodes,
  nodes={inner sep=0pt},
  column sep=15pt,
  row sep=6pt,
  ampersand replacement=\&,
  below=60pt of m1.south
] {
  \tikz\pic{person={cfOrange}{45pt}{0}}; \&
  \tikz\pic{person={cfOrange}{45pt}{0}}; \&
  \tikz\pic{person={cfOrange}{45pt}{0}}; \\
  \tikz\pic{person={cfOrange}{45pt}{0}}; \&
  \tikz\pic{person={cfOrange}{45pt}{0}}; \&
  \tikz\pic{person={cfOrange}{45pt}{0}}; \\
  \tikz\pic{person={cfOrange}{45pt}{0}}; \&
  \tikz\pic{person={cfOrange}{45pt}{0}}; \&
  \tikz\pic{person={cfOrange}{45pt}{0}}; \\
};

\matrix (m4) [
  matrix of nodes,
  nodes={inner sep=0pt},
  column sep=15pt,
  row sep=6pt,
  ampersand replacement=\&,
  right=45pt of m3.east
] {
  \tikz\pic{person={MediumBlue}{45pt}{0}}; \&
  \tikz\pic{person={MediumBlue}{45pt}{0}}; \&
  \tikz\pic{person={MediumBlue}{45pt}{0}}; \\
  \tikz\pic{person={MediumBlue}{45pt}{0}}; \&
  \tikz\pic{person={MediumBlue}{45pt}{0}}; \&
  \tikz\pic{person={MediumBlue}{45pt}{0}}; \\
  \tikz\pic{person={MediumBlue}{45pt}{0}}; \&
  \tikz\pic{person={MediumBlue}{45pt}{0}}; \&
  \tikz\pic{person={MediumBlue}{45pt}{0}}; \\
};
\draw[
  rounded corners=8pt,
  very thick
] ($(m1.north west) + (-4pt,4pt)$) 
rectangle 
($(m1.south east) + (4pt,-4pt)$)
node[midway, below=77pt, font=\bfseries, align = center] {Treatment};
\draw[thick, gray!75]
  ($(m1-3-1.north west) + (-2.5pt,2.5pt)$)
  rectangle
  ($(m1-3-1.south east) + (2.5pt,-2.5pt)$);

\draw[->, thick, dashed, gray!75]
  ($(m1-1-1.south west) + (3pt,2pt)$)
  to[bend right=15]
  ($(m1-3-1.north west) + (-1pt,3.5pt)$);
\draw[->, thick, dashed, gray!75]
  ($(m1-2-1.west) + (1.5pt,-9pt)$)
  to[bend right=20]
  ($(m1-3-1.north west) + (3pt,3.5pt)$); 
\draw[->, thick, dashed, gray!75]
  ($(m1-3-2.west) + (3pt,-14pt)$)
  to[bend left=15]
  ($(m1-3-1.east) + (3.5pt,-11pt)$);  
\draw[->, thick, dashed, gray!75]
  ($(m1-3-3.south west) + (2.5pt,3.5pt)$)
  to[bend left=25]
  ($(m1-3-1.south east) + (3.5pt,2pt)$);
\draw[->, thick, dashed, gray!75]
  ($(m1-2-2.south west) + (2.5pt,1pt)$)
  to[bend left=25]
  ($(m1-3-1.north east) + (3.5pt,-2pt)$);
\draw[thick, dashed, gray!75]
  ($(m1-2-3.south west) + (2.5pt,1pt)$)
  to[bend left=20]
  ($(m1-3-2.north east) + (-1.5pt,-13pt)$);
\draw[->, thick, dashed, gray!75]
  ($(m1-3-2.west) + (-1pt,4pt)$)
  to[bend left=10]
  ($(m1-3-1.east) + (3.5pt,3.5pt)$);
\draw[->, thick, dashed, gray!75]
  ($(m1-1-2.south west) + (3pt,2pt)$)
  to[bend right=12]
  ($(m1-3-1.north east) + (-1pt,3.5pt)$);
\draw[thick, dashed, gray!75]
  ($(m1-1-3.south west) + (3pt,2pt)$)
  to[bend right=12]
  ($(m1-2-2.north) + (7pt,-3pt)$);
\draw[->, thick, dashed, gray!75]
  ($(m1-2-2.west) + (-1pt,1pt)$)
  to[bend right=12]
  ($(m1-3-1.north east) + (2pt,3.5pt)$);  

\draw[->, thick, dashed, gray!75]
  ($(m2-1-1.south west) + (3pt,2pt)$)
  to[bend right=15]
  ($(m2-3-1.north west) + (-1pt,3.5pt)$);
\draw[->, thick, dashed, gray!75]
  ($(m2-2-1.west) + (1.5pt,-9pt)$)
  to[bend right=20]
  ($(m2-3-1.north west) + (3pt,3.5pt)$); 
\draw[->, thick, dashed, gray!75]
  ($(m2-3-2.west) + (3pt,-14pt)$)
  to[bend left=15]
  ($(m2-3-1.east) + (3.5pt,-11pt)$);  
\draw[->, thick, dashed, gray!75]
  ($(m2-3-3.south west) + (2.5pt,3.5pt)$)
  to[bend left=25]
  ($(m2-3-1.south east) + (3.5pt,2pt)$);
\draw[->, thick, dashed, gray!75]
  ($(m2-2-2.south west) + (2.5pt,1pt)$)
  to[bend left=25]
  ($(m2-3-1.north east) + (3.5pt,-2pt)$);
\draw[thick, dashed, gray!75]
  ($(m2-2-3.south west) + (2.5pt,1pt)$)
  to[bend left=20]
  ($(m2-3-2.north east) + (-1.5pt,-13pt)$);
\draw[->, thick, dashed, gray!75]
  ($(m2-3-2.west) + (-1pt,4pt)$)
  to[bend left=10]
  ($(m2-3-1.east) + (3.5pt,3.5pt)$);
\draw[->, thick, dashed, gray!75]
  ($(m2-1-2.south west) + (3pt,2pt)$)
  to[bend right=12]
  ($(m2-3-1.north east) + (-1pt,3.5pt)$);
\draw[thick, dashed, gray!75]
  ($(m2-1-3.south west) + (3pt,2pt)$)
  to[bend right=12]
  ($(m2-2-2.north) + (7pt,-3pt)$);
\draw[->, thick, dashed, gray!75]
  ($(m2-2-2.west) + (-1pt,1pt)$)
  to[bend right=12]
  ($(m2-3-1.north east) + (2pt,3.5pt)$);  
  
\draw[
  rounded corners=8pt,
  very thick
] ($(m2.north west) + (-4pt,4pt)$) 
rectangle 
($(m2.south east) + (4pt,-4pt)$)
node[midway, below=77pt, font=\bfseries, align = center] {Control};
\draw[thick, gray!75]
  ($(m2-3-1.north west) + (-2.5pt,2.5pt)$)
  rectangle
  ($(m2-3-1.south east) + (2.5pt,-2.5pt)$);

\draw[
  rounded corners=8pt,
  very thick
] ($(m3.north west) + (-4pt,4pt)$) 
rectangle 
($(m3.south east) + (4pt,-4pt)$)
node[midway, below=77pt, font=\bfseries, align = center] {Treatment};
\draw[thick, gray!75]
  ($(m3-3-1.north west) + (-2.5pt,2.5pt)$)
  rectangle
  ($(m3-3-1.south east) + (2.5pt,-2.5pt)$);

\draw[->, thick, dashed, gray!75]
  ($(m3-1-1.south west) + (3pt,2pt)$)
  to[bend right=15]
  ($(m3-3-1.north west) + (-1pt,3.5pt)$);
\draw[->, thick, dashed, gray!75]
  ($(m3-2-1.west) + (1.5pt,-9pt)$)
  to[bend right=20]
  ($(m3-3-1.north west) + (3pt,3.5pt)$); 
\draw[->, thick, dashed, gray!75]
  ($(m3-3-2.west) + (3pt,-14pt)$)
  to[bend left=15]
  ($(m3-3-1.east) + (3.5pt,-11pt)$);  
\draw[->, thick, dashed, gray!75]
  ($(m3-3-3.south west) + (2.5pt,3.5pt)$)
  to[bend left=25]
  ($(m3-3-1.south east) + (3.5pt,2pt)$);
\draw[->, thick, dashed, gray!75]
  ($(m3-2-2.south west) + (2.5pt,1pt)$)
  to[bend left=25]
  ($(m3-3-1.north east) + (3.5pt,-2pt)$);
\draw[thick, dashed, gray!75]
  ($(m3-2-3.south west) + (2.5pt,1pt)$)
  to[bend left=20]
  ($(m3-3-2.north east) + (-1.5pt,-13pt)$);
\draw[->, thick, dashed, gray!75]
  ($(m3-3-2.west) + (-1pt,4pt)$)
  to[bend left=10]
  ($(m3-3-1.east) + (3.5pt,3.5pt)$);
\draw[->, thick, dashed, gray!75]
  ($(m3-1-2.south west) + (3pt,2pt)$)
  to[bend right=12]
  ($(m3-3-1.north east) + (-1pt,3.5pt)$);
\draw[thick, dashed, gray!75]
  ($(m3-1-3.south west) + (3pt,2pt)$)
  to[bend right=12]
  ($(m3-2-2.north) + (7pt,-3pt)$);
\draw[->, thick, dashed, gray!75]
  ($(m3-2-2.west) + (-1pt,1pt)$)
  to[bend right=12]
  ($(m3-3-1.north east) + (2pt,3.5pt)$);  

\draw[->, thick, dashed, gray!75]
  ($(m4-1-1.south west) + (3pt,2pt)$)
  to[bend right=15]
  ($(m4-3-1.north west) + (-1pt,3.5pt)$);
\draw[->, thick, dashed, gray!75]
  ($(m4-2-1.west) + (1.5pt,-9pt)$)
  to[bend right=20]
  ($(m4-3-1.north west) + (3pt,3.5pt)$); 
\draw[->, thick, dashed, gray!75]
  ($(m4-3-2.west) + (3pt,-14pt)$)
  to[bend left=15]
  ($(m4-3-1.east) + (3.5pt,-11pt)$);  
\draw[->, thick, dashed, gray!75]
  ($(m4-3-3.south west) + (2.5pt,3.5pt)$)
  to[bend left=25]
  ($(m4-3-1.south east) + (3.5pt,2pt)$);
\draw[->, thick, dashed, gray!75]
  ($(m4-2-2.south west) + (2.5pt,1pt)$)
  to[bend left=25]
  ($(m4-3-1.north east) + (3.5pt,-2pt)$);
\draw[thick, dashed, gray!75]
  ($(m4-2-3.south west) + (2.5pt,1pt)$)
  to[bend left=20]
  ($(m4-3-2.north east) + (-1.5pt,-13pt)$);
\draw[->, thick, dashed, gray!75]
  ($(m4-3-2.west) + (-1pt,4pt)$)
  to[bend left=10]
  ($(m4-3-1.east) + (3.5pt,3.5pt)$);
\draw[->, thick, dashed, gray!75]
  ($(m4-1-2.south west) + (3pt,2pt)$)
  to[bend right=12]
  ($(m4-3-1.north east) + (-1pt,3.5pt)$);
\draw[thick, dashed, gray!75]
  ($(m4-1-3.south west) + (3pt,2pt)$)
  to[bend right=12]
  ($(m4-2-2.north) + (7pt,-3pt)$);
\draw[->, thick, dashed, gray!75]
  ($(m4-2-2.west) + (-1pt,1pt)$)
  to[bend right=12]
  ($(m4-3-1.north east) + (2pt,3.5pt)$);  
  
\draw[
  rounded corners=8pt,
  very thick
] ($(m4.north west) + (-4pt,4pt)$) 
rectangle 
($(m4.south east) + (4pt,-4pt)$)
node[midway, below=77pt, font=\bfseries, align = center] {Control};
\draw[thick, gray!75]
  ($(m4-3-1.north west) + (-2.5pt,2.5pt)$)
  rectangle
  ($(m4-3-1.south east) + (2.5pt,-2.5pt)$);  

\node at ($(m1)!0.5!(m2)$) {vs.};
\node at ($(m3)!0.5!(m4)$) {vs.};

\node[above=10pt of m1.north west, anchor=south west, xshift=-30pt] {\small a) Person-Level Causal Contrast in Local Intervention};
\node[above=10pt of m3.north west, anchor=south west, xshift=-30pt] {\small b) Person-Level Causal Contrast in Global Intervention};

\coordinate (legendSW) at ($(m3.south west)+(33pt,-40pt)$);
\coordinate (legendSE) at ($(m4.south east)+(33pt,-40pt)$);

\node[draw, color = cfOrange, fill=cfOrange, minimum size=7pt, anchor=west] (leg1) at (legendSW) {};
\node[right=0.1cm of leg1, anchor=west] {\scriptsize Encouraged to quit social media};

\node[draw, color = MediumBlue, fill=MediumBlue, minimum size=7pt, anchor=west, below=0.1cm of leg1] (leg2) {};
\node[right=0.1cm of leg2, anchor=west] {\scriptsize Not encouraged to quit social media};

\node[anchor=west, below=0cm of leg2,xshift=79pt] (leg3)
{
  \tikz[baseline=-0.5ex]{\draw[dashed,gray,->, thick] (0,0) -- (0.3cm,0);} \hspace{-2pt} \scriptsize A's treatment status affects B's outcome
};

\node[fit=(leg1) (leg3), draw, inner xsep=6pt, inner ysep=6pt, thick] (legendbox) {};
  
\end{tikzpicture}
}
\caption{\small \textbf{Local- and global-intervention effects involve different comparisons of potential outcomes.} Each black rectangle represents the same cluster of students under a different intervention. Light gray boxes highlight a focal student in that cluster. Because other students' treatment assignments affect the focal student's outcome, her individual treatment effect plausibly differs between the two interventions.}
\label{fig:local}
\end{figure}

If person $i$'s mental health is affected by another population member's treatment status, $Y_i(Z_i)$ is ill-defined. Her depressive affect will vary as a function of whether person $j$ or person $k$ is encouraged to quit social media (for $i \neq j \neq k$)---a phenomenon known as \textit{interference}. We can more precisely define person $i$'s potential outcome using the treatment statuses of all members of the population. Let $\bm{Z}$ represent the vector of treatment statuses for the entire population. Let $\bm{0}_{n}$ denote a vector of length $n$ containing only 0s: $\bm{0}_{n} = (0_{1}, 0_{2},\ldots,0_{n})$. Define $\bm{1}_{n}$ analogously. $Y_i(\bm{Z} = \bm{1}_{n})$ represents person $i$'s outcome had every member of the population been encouraged to quit social media, and $Y_i(\bm{Z} = \bm{0}_{n})$ her potential outcome had no members been encouraged. Person $i$'s global-intervention effect is therefore given by $Y_i(\bm{Z} = \bm{1}_{n}) - Y_i(\bm{Z} = \bm{0}_{n})$. We can now define a population average global-intervention effect.

\begin{definition}
Average Global-Intervention Effect (AGIE).

\begin{equation}
\tau_{\text{AGIE}} \equiv 
\begingroup
\color[HTML]{237023}
\underbrace{
\color{black}\frac{1}{n}\sum^n_{i=1} 
\begingroup
\color[HTML]{377eb8}
\underbrace{\hspace{0.5em}\color{black}Y_i(\bm{Z} = \bm{1}_n)\hspace{0.5em}}_{\mathclap{\substack{\text{Person $i$'s} \\ \text{depressive affect} \\ \text{if everybody,} \\ \text{including her,} \\ \text{is encouraged to} \\ \text{quit social media...}}}}
\endgroup
\begingroup
\color[HTML]{377eb8}
\underbrace{
\hspace{0.5em}\color{black} -
\quad
Y_i(\bm{Z} = \bm{0}_{n}).}_{\mathclap{\substack{\text{...minus the same} \\ \text{person's depressive } \\ \text{affect if nobody,} \\ \text{ including her,} \\ \text{is encouraged to} \\ \text{quit social media...}}}}
\endgroup
\hspace{0.25em}}_{\mathclap{\substack{\text{...averaged across every person in the population.}}}}
\endgroup
\end{equation}
\end{definition}

The same estimand has been described as a \textit{global average treatment effect}, a \textit{general-equilibrium effect}, and an \textit{all-or-nothing effect} \citep{heckman1,eate,molly}.\footnote{I avoid the term \textit{global average treatment effect} because the name \textit{local average treatment effect} refers to a different causal estimand in which \textit{local} refers to the sub-population across which effects are averaged rather than the scale of the intervention \citep{AIR96}. The compound adjective \textit{global-intervention} serves to clarify that it is the intervention, not the average, that is global.} The AGIE can also be seen as a special case of the \textit{overall effect} as defined in the interference literature \citep{vand1,vand2}.

To define local-intervention effects, let $\bm{Z}_{-i}$ represent the treatment assignment vector for everybody except person $i$. For instance, $\bm{Z}_{-i} = \bm{1}_{n-1}$ indicates that, with the possible exception of person $i$, everybody in the population was encouraged to quit social media. Person $i$'s local-intervention effect can be defined as $Y_i(Z_i = 1, \bm{Z}_{-i} = \bm{0}_{n-1}) - Y_i(Z_i = 0, \bm{Z}_{-i} = \bm{0}_{n-1})$. This is the effect of encouraging person $i$, but no one else in the entire population, to quit social media. We can now define a population average local-intervention effect.

\begin{definition}
Average Local-Intervention Effect (ALIE).

\begin{equation}
\tau_{\text{ALIE}} \equiv 
\begingroup
\color[HTML]{237023}
\underbrace{
\color{black}\frac{1}{n}\sum^n_{i=1}
\begingroup
\color[HTML]{377eb8}
\underbrace{\color{black}\hspace{0.25em}Y_i(Z_i = 1, \bm{Z}_{-i} = \bm{0}_{n-1})\hspace{0.25em}}_{\mathclap{\substack{\text{Person $i$'s depressive affect} \\ \text{if she, but no one else,} \\ \text{is encouraged to} \\ \text{quit social media...}}}} 
\endgroup
\begingroup
\color[HTML]{377eb8}
\underbrace{\color{black} \hspace{0.25em}- \hspace{0.5em} Y_i(Z_i = 0, \bm{Z}_{-i} = \bm{0}_{n-1}).}_{\mathclap{\substack{\text{...minus the same person's depressive } \\ \text{affect if nobody, including} \\ \text{her, is encouraged to} \\ \text{quit social media...}}}} 
\endgroup
\hspace{0.25em}}_{\mathclap{\substack{\text{...averaged across every person in the population.}}}}
\endgroup
\end{equation}
\end{definition}

\noindent As I've defined it, the ALIE is a special case of what is called the \textit{direct effect} in \citet{vand1} and \citet{vand2}.\footnote{The definitions in these references differ slightly from the definitions in \citep{hudgens}. I avoid the terms \textit{direct effect} and \textit{overall effect} because \textit{direct} is widely used to in the mediation literature in the names of several different estimands.} \autoref{fig:local} visually depicts the difference between person-level local-intervention effects and person-level global-intervention effects and may help readers better understand Equations 1 and 2.\footnote{In the causal inference lexicon, a \textit{person-level effect} or \textit{individual treatment effect} is the effect of a treatment for a specific person. This is why I avoid the labels ``person-level'' and ``group-level,'' as used in \citet{group1} and \citet{group2}, to describe local- and global-intervention effects.\label{fn:group}}

Within the population of high schoolers, no restrictions on interference have been invoked. Under a partial interference assumption, however, we can offer an alternative expression of the ALIE that better illustrates the connection to cluster-randomized trials. Suppose high schools are indexed by $g = 1,2,\ldots,m$, with the number of students within each high school indexed by $n_g$, with $n_g \cdot m = n$. Assume that interference does not occur across high schools but may occur within a school. For instance, if Alice and Bill attend the same school, Alice's treatment status may affect Bill's mental health. If Carol attends a different school, however, her treatment status must have no effect on Alice's mental health or Bill's. Each student's potential outcome can then be written as a function of her treatment assignment and the vector of treatment assignments for her schoolmates only. Let $Y_{g,i}$ represent the outcome for person $i$ in high school $g$, and let $\bm{Z}_{g,-i}$ represent the assignment vector for every student in high school $g$ except student $i$. We can now offer a different representation of the ALIE, although it is equivalent to Definition 2 under the partial interference assumption described above.\footnote{We could also define a version of the ALIE in which each cluster, rather than each person, receives equal weight.}

\begin{align}
\begin{split}
\tau_{\text{ALIE}} = 
\frac{1}{n}\sum^m_{g=1}\sum_{i \in g}
\hspace{0.25em}Y_{g,i}(&Z_{g,i} = 1, \bm{Z}_{g,-i} = \bm{0}_{n_{g}-1})\\
&- Y_{g,i}(Z_{g,i} = 0, \bm{Z}_{g,-i} = \bm{0}_{n_{g}-1}).
\end{split}
\end{align}

\noindent An analogous representation of the AGIE can also be given by forcing $\bm{Z}_{g,-i} = \bm{1}_{n_{g}-1}$ in the first potential outcome. 

Partial interference assumptions like the above are common in the contagion literature and enable the identification of the AGIE and the ALIE \citep{vand1}. Alternatively, researchers may allow interference to extend beyond a given cluster but assume it declines as units become more socially distant \citep{leung1,leung2,linear}. Such \textit{spillover decay} assumptions can enable consistent but potentially biased estimation of global-intervention effects. In the traditional finite-population framework, under limited interference but otherwise arbitrary and unknown interference, individually-randomized trials can identify the expected average treatment effect (EATE) of \citet{eate}, where the expectation is taken across all treatment assignment vectors that could have been generated by treatment randomization.\footnote{Interference must be limited in the average number of \textit{interference dependencies} per participant. An interference dependency occurs whenever two participants are affected by the same participant's treatment status. This may occur when the two participants interfere with each other or when both participants are affected by the treatment status of a third participant. See \citet[][p. 681]{eate} for further discussion of interference dependencies.} For instance, in a balanced completely randomized experiment, in which the number of participants assigned treatment is fixed at half the sample size, the EATE would only marginalize across assignment vectors with exactly half of all units assigned treatment. The ALIE, in contrast, fixes the assignments at 0 for all but one participant (or, under partial interference, for all but one participant in a given cluster). 

\subsubsection{Alternative Estimands}\label{sec:alt}

Slight variations on the AGIE and ALIE, as I define them, may also be of interest to researchers. First, we might define the AGIE and ALIE in terms of treatment status (actual social media) rather than treatment assignment (encouragement to quit social media). Second, if defined in terms of treatment status, we might be interested in the effects of nearly-global interventions in which, say, 90\% of social media users quit. Such an estimand would, like the global-intervention effect, represent a special case of the overall effect \citep{vand1,vand2}.\footnote{Note that such estimands are easier to interpret when it only matters \textit{how many} people are treated and not \textit{which} people are treated, which is sometimes called a \textit{stratified interference} assumption. Defining global- and local-intervention effects in terms of treating nobody but person $i$ or everybody including person $i$ helped us avoid this complication.} We might also consider interventions that are small in scale but highly targeted. For instance, we can imagine assigning Alice and her very closest friends to treatment while assigning Bill and his very closest friends to control. If we believe close friend groups contain the strongest spillover effects, then such targeted, small-scale interventions might prove almost as effective as large-scale interventions. Two-stage randomized trials, in which clusters are randomly assigned to receive a particular treatment program and then participants within a cluster are randomly assigned to the corresponding treatments, could enable the identification of these estimands as well.

\subsection{A Subtlety Regarding No-Interference Assumptions}\label{sec:interference}

The most commonly invoked restriction on interference is the no-interference assumption, which requires each person's potential outcome to be a function of only her treatment assignment.\footnote{Following \cite{vand3}, I separate the no-interference assumption from the broader stable-unit-treatment-value assumption (SUTVA). The no-interference assumption constitutes only part of SUTVA.} Although applied researchers seldom state it explicitly, the standard average treatment effect is ill-defined without it.

It is tempting to conclude that this assumption is violated in most published RCTs on social media use. But that conclusion, as stated, is imprecise. The reason is that there are two different versions of the no-interference assumption. In the \textit{super-population} framework, the assumption prohibits interference between all members of the population from which the trial participants are drawn. In the \textit{finite-population} framework, however, the assumption prohibits interference only between trial participants, as the participants constitute the entire population of interest in this scheme. Importantly, it is possible for the finite-population version to hold even when the super-population version fails.

To see why, return to \textit{Local Quitting Trial}. In this hypothetical trial, we restricted interference to occur only within schools. Suppose further that researchers recruit one student each from a set of schools to conduct an RCT on quitting social media. This design ensures no interference occurs between trial participants: the social media use of one participant has no effect on the mental health of any other participants. But the social media use of non-participants may still affect the mental health of participants. For instance, suppose Alice and Bill attend the same school and Alice (but not Bill) participates in the trial. Bill's social media use may still affect Alice's mental health. In this scenario, finite-population non-interference holds, but the super-population version does not.

To formalize this idea, I return to the notational scheme used in the previous section. Assume researchers draw a sample from the population that is smaller than the entire population. Let $i \in \mathcal{S}$ indicate that person $i$ is in the $\mathcal{S}$ample. Let $\bm{Z}_{\mathcal{S}\backslash\{i\}}$ represent the treatment assignment vector for all trial participants except person $i$, and let $\bm{Z}_{\mathcal{S}^c}$ represent the assignment vector for all members of the population except those who were sampled. We can then define both a sample non-interference assumption and a population non-interference assumption.

\begin{definition}
Sample Non-Interference.

\begin{equation}
Y_i(z_i, \bm{z}_{\mathcal{S}^c}, \bm{z}_{\mathcal{S}\backslash\{i\}}) = Y_i(z_i, \bm{z}_{\mathcal{S}^c}, \bm{z'}_{\mathcal{S}\backslash\{i\}})
\hspace{.5em}\forall\hspace{.3cm} \bm{z}_{\mathcal{S}^c},\bm{z}_{\mathcal{S}\backslash\{i\}},\bm{z'}_{\mathcal{S}\backslash\{i\}};\hspace{.3cm}\forall\hspace{.3cm}i \in \mathcal{S}.
\end{equation}
\end{definition}

\noindent Sample non-interference ensures that a given participant's potential outcome is not a function of any other participant's assignment. The treatment assignments of people outside the trial, however, may still affect outcomes of trial participants. (Treatment assignment for every non-participant is of course 0, but this is not necessarily the case for treatment status.) Next, we can define a population non-interference assumption.

\begin{definition}
Population Non-Interference.

\begin{equation}
Y_i(z_i, \bm{z}_{-i}) = Y_i(z_i, \bm{z'}_{-i}) \hspace{.3cm}\forall\hspace{.3cm} \bm{z}_{-i},\bm{z'}_{-i};
\hspace{.3cm}\forall\hspace{.3cm}i.
\end{equation}

\end{definition}

\noindent Within this framework, it possible for sample non-interference to hold while population non-interference is violated. 

\section{Further Remarks}\label{sec:further}

In this section, I discuss two other complications related to the social media hypothesis estimand and clarify the scope of my argument concerning local- and global-intervention effects.

\subsection{The Social Media Hypothesis and Multiple Versions of Treatment}\label{sec:compound}

For simplicity, I have described the social media hypothesis solely in terms of social media use. Two complications are worth mentioning.

The first is that ``using social media'' is a treatment with multiple \textit{versions}, in the causal inference lexicon. There are different platforms, and each can be used in any number of ways.\footnote{A further source of ``hidden versions'' of social media use is that social media platforms have changed substantially over time, becoming more image-based and incorporating more features designed to boost time spent on the platforms \citep{image,addict}.} These versions may have different effect sizes: the effect of using Facebook for five minutes a day may differ from the effect of using Instagram for five hours a day. It might be argued, then, that the treatment ``using social media'' is ill-defined and that any treatment effect of social media use is therefore ill-defined. But this is too strong a stance. It is rarely feasible to eliminate hidden versions of treatment entirely, and the estimand identified in such studies can still be clearly defined.\footnote{See \citet{precise} and \citet{vandvague} for further discussion on the inherent vagueness of treatment definitions.} In particular, \citet{hiddenestimand} show that the estimand can be interpreted as a weighted average of different version-specific treatment effects, with weights proportional to a version's prevalence in the study population. Descriptive research on how people typically use social media can therefore inform how we interpret the effects identified by social media RCTs.

The second complication concerns the role of smartphones in the decline of teen mental health. As mentioned, proponents of the social media hypothesis typically attribute much of the decline in teen mental health to the widespread adoption of both smartphones and social media \citep{smh1,smh2}. We might consider an alternative formulation of the social media hypothesis that attributes the decline to excessive use of social media on smartphones in particular, or to excessive use of both social media and smartphones. In other words, the social media hypothesis can also be formulated in terms of a \textit{composite}, or \textit{compound}, intervention: the widespread adoption of both social media and smartphones. Indeed, the widespread adoption of smartphones helped transform social media by enabling the growth of image- and video-based platforms. I avoid this complication for simplicity, but it is another issue to keep in mind when interpreting the effects of RCTs that reduce social media use but not smartphone use.

\subsection{The Social Media Hypothesis and Causal Attribution}\label{sec:attribution}

Recall that the social media hypothesis attributes a large share of the decline in teen mental health to the widespread adoption of social media. It is a claim about the causes of a particular effect and therefore represents what is called a \textit{causes-of-effects} estimand, or causal attribution estimand. RCTs, along with most causal observational studies, identify the effects of a particular cause---\textit{effects-of-causes} estimands \citep{coe1,coe2}. 

To understand the distinction, it helps to consider a simpler example. Suppose we run a trial that randomly assigns a particular surgical intervention to patients with heart disease. We can ask an effects-of-causes question about this intervention: how much does the surgery increase the probability of surviving until the end of follow-up? But we can also ask a causes-of-effects question. Suppose that Alice participates in the trial, receives the surgery, and survives until the end of follow-up. Did Alice survive \textit{because} she received the surgery, or would she have survived even without it? The social media hypothesis is a type of causes-of-effects claim: it attributes something that already happened to a particular intervention. 

No work, to my knowledge, has formalized the identification of causal attribution effects for demographic trends like the decline in teen mental health. Some work has focused on identifying the probability that a particular factor caused a discrete event to occur, such as a war breaking out or a specific person's death. \citet{pearl} uses a monotonicity assumption to identify the probability that a particular treatment caused a discrete outcome (see also \citet{rosenbaum}). In the present setting, monotonicity would require that social media use has no positive treatment effect on mental health for any member of the population. This is an exceedingly strong assumption unlikely to hold in practice. More recently, \citet{stratton} show how to identify the proportion of cross-sectional population variance causally explained by the observed variation in a treatment without a monotonicity assumption, but it is unclear how to incorporate global-intervention effects into such an approach. It is also unclear what it would mean for such an approach if, say, the effects of social media use varied over time. For the time being, causal attribution effects for demographic trends may be best assessed informally. Proponents of the social media hypothesis have offered such informal assessments by attempting to rule out alternative explanations for the decline in teen mental health, although this exercise faces its own methodological challenges \citep{competing}.\footnote{One such challenge is that some ``treatments''---economic transformation, widespread adoption of social media---unfold gradually over time. It is unclear, a priori, when the effects of such gradually unfolding processes should start to appear. Another complication is that two causal factors can potentially interact, each enhancing the other's effects on mental health.}

A final point about causal attribution is worth making. It is tempting to dismiss causal attribution questions as policy-irrelevant: we cannot change the past, so why worry about the causes of past events? But this dismissal is too quick. Even if causal attribution claims are about the past, they can offer indirect evidence about potential future policies. If global-intervention effects are difficult to credibly identify, ruling out competing explanations for the decline in teen mental health may make us more confident that global social media interventions have large effects. 

\subsection{Observational Studies On Social Media Use}\label{sec:obs}

I have focused on social media RCTs for two reasons. First, RCT estimates are typically seen as more credible than results from observational studies. Second, hypothetical trials are a useful tool for precisely describing causal estimands without using mathematical notation. But the points I have raised about intervention scale, intervention type, and intervention duration apply to many published observational studies as well. 

A review of observational studies on social media harms is beyond the scope of this paper. For illustration, however, consider the Monitoring the Future (MtF) dataset, which has been used in some of the most highly cited studies on social media use, including both studies that report ``large'' effects and studies that report ``small'' effects \citep{OP,twengeOP,twengeOP2}. The MtF data is cross-sectional and contains no information on past social media use, making it impossible to determine a student's history of social media use. Moreover, interference is likely substantial: the data is collected by first randomly selecting schools and then surveying either every available student or a large portion of the available students in each school, with no school identification variables present in the dataset \citep{MtF}. Researchers have no way to distinguish between different clusters of students. Even in the absence of unmeasured confounding and reverse causation, such studies would at best approximate an average of local-intervention effects with an unknown mix of treatment histories.

More generally, my remarks on the shortcomings of social media RCTs should not be taken as a general critique of RCTs or a call for fewer RCTs and more observational studies. RCTs provide epistemic guarantees that observational studies cannot, and generating experimental evidence plays a critical role in advancing social scientific research. But whether the study is observational or experimental, we must think carefully about what estimand is most plausibly identified and what that estimand actually tells us about the social world.

\subsection{Policy-Relevant Social Media Estimands}

I have framed my discussion in terms of the social media hypothesis, but it could also be framed in terms of the policy relevance of the RCT estimand. Public policy could prevent many young adolescents from joining social media until they are much older; the long-term global joining effect is more policy relevant than the short-term local quitting effect.

\section{Learning About Global-Intervention Effects From Imperfect Evidence}\label{sec:conclusion}

In this paper, I have pointed out that the estimand identified by social media RCTs differs substantially from the estimand invoked by the social media hypothesis. In other words, extant social media RCTs fail to offer an appropriate test of the social media hypothesis; an ideal trial designed to assess the hypothesis would drastically differ from the trials that have actually been carried out. I have devoted special attention to the difference in intervention scale, with individually-randomized trials identifying effects of local interventions and the social media hypothesis positing large effects of a global intervention. To conclude, I offer suggestions on how researchers might learn more about long-term global social media interventions in the absence of perfect identification.

Cluster-randomized trials would offer some of the strongest evidence about large-scale social media interventions, and one recent cluster-randomized trial provides suggestive evidence that large-scale social media interventions yield larger per-person effects \citep{kuipers}. Cluster-randomized trials, however, typically face serious methodological obstacles. Cross-cluster interference may occur, and attrition may be high. Point identification of intent-to-treat effects under such complications relies on strong assumptions that are not guaranteed to hold. Moreover, non-compliance renders intent-to-treat effects less informative. Non-compliance may be substantial in cluster-randomized trials: \citet{kuipers} note that among non-focal cluster members, compliance with deactivation was around 33\%. Partial identification of global treatment effects among compliers is still possible in such settings, but point identification is not \citep{keele1}.

Observational studies on social media effects can be informative as well. School-level smartphone bans and country-level age limits on social media use present opportunities for scholars to learn more about large-scale intervention effects through research designs like the synthetic control method, difference-in-differences, or interrupted time series \citep{SC1,SC2,did1,did2,coe2}. These strategies are not without their problems. Each approach relies on strong assumptions that are most credible for studying short-term rather than long-term treatment effects \citep{dangers}. Moreover, school-wide smartphone bans do not prohibit social media use outside of school, and limiting social media to those older than 16 would not eliminate spillover effects for those 16 and under. Nonetheless, if intervention scale is important, we should still expect effects from such cluster-level studies to be larger than those estimated in analogous person-level studies.

Descriptive research---both quantitative and qualitative---may prove especially useful for learning about global-intervention effects. Surveys, interviews, and ethnographies could inform us about the prevalence of content sharing or social exclusion as described in \autoref{sec:why} \citep{collectiveaction}. Ethnographic research could help paint a richer portrait of social media--era teenage life and offer deeper insights into how social media's widespread adoption may have altered youth culture. 

Cluster-randomized trials, cluster-level observational studies, and descriptive work each suffer from their own shortcomings. Each piece of evidence would, on its own, provide only imperfect evidence about long-term global-intervention effects. But triangulating these difference pieces of evidence could help us learn more about the effects of sustained, widespread social media use \citep{haack,levimartin}. Synthesizing various forms of evidence has played an important role in advancing scientific knowledge, notably in work on cigarettes and lung cancer \citep{cornfield}. How exactly researchers should carry out this activity is beyond the scope of this article, but this kind of synthetic work may prove to be critical to ongoing debates about teen mental health.\footnote{It should be emphasized that this kind of synthetic work should not simply combine different studies that all suffer from the same methodological shortcoming. For instance, if we are suspicious of cross-sectional observational studies primarily due to the threat of reverse causation, simply combining results from more cross-sectional observational studies fails to address the central problem. What would be needed in such a case is a longitudinal study that obviates concerns of reverse causation (but perhaps suffers from other methodological problems, which could in turn be addressed in additional work, and so on).} It still remains crucial for authors of original research to carefully specify estimands of interest and the assumptions sufficient for identifying such estimands. But given that identification assumptions are almost always suspect, bringing together different sources of evidence may be the best way to learn about global-intervention effects in social media research.

\newpage

\bibliography{SMR.refs}

\newpage
\begin{appendices}

\end{appendices}

\end{document}